\documentclass[11pt,longbibliography,superscriptaddress]{revtex4-2}

\usepackage[utf8]{inputenc}
\usepackage{amsmath}
\usepackage{amssymb}
\usepackage{amsmath}
\usepackage{appendix}
\usepackage{bbold}
\usepackage{color}

\begin{document}
\title{Causality in  String  Field Theory}

\author{Arshid Shabir }
\affiliation{Department of Physics, NIT Srinagar, Kashmir 190006 India}

\author{Naveed Ahmad Shah}
 		\affiliation{Department of Physics, Jamia Millia Islamia, New Delhi  110025, India}

\author{Salman Sajad Wani}
	\affiliation{Department of Physics Engineering, Istanbul Technical University, Istanbul, 34469  Turkey}

\author{Mir Faizal}
\affiliation{Canadian Quantum Research Center, 204-3002, 32 Ave Vernon, BC V1T 2L7 Canada}
\affiliation{Irving K. Barber School of Arts and Sciences, University of British Columbia Okanagan, Kelowna, BC V1V 1V7, Canada} 

 \author{Kousar Jan }
\affiliation{Department of Physics, NIT Srinagar, Kashmir 190006 India}
\author{Seemin Rubab}
\affiliation{Department of Physics, NIT Srinagar, Kashmir 190006 India}

\begin{abstract}
In this letter, we will investigate causality in string field theory using  pp-wave  light-cone gauge string field theory. We will generalize the Ramsey scheme to string field theory, and use it to  analyze  string field theoretical processes.   An explicit characteristic function for interactive string field theory will be built using this string field theoretical Ramsey scheme. The average of the difference between the initial and final values of any operator described in string field theory will be obtained using this characteristic function.  We will use the quantum information theoretical technique based on quantum fisher information to extract information about such  string field theoretical processes. 
\end{abstract}

\maketitle
\section{Introduction}

The  string field theory is the second quantized version of string theory \cite{witten}. As it is a consistent second quantized version of string theory, it can be used to analyze   off-shell string amplitude \cite{shell1, shell2}.
Furthermore, string field  theory is background independent  \cite{back1, back2}. It has been argued that the  divergences    due to a Euclidean world-sheet description for the boundary of moduli space can be removed  in string field theory \cite{sft01, sft02}.  The problems relating to the unitarity of string theory are also revolved in  string field theory   \cite{prob6}. String field theory can be used to address the  crossing symmetry of scattering amplitudes in string theory \cite{prob7}.   It is possible to construct  string field theory  in  light-cone gauge \cite{light}. In fact, it is possible to relate the covariant string field theory to the light-cone gauge string field theory \cite{light2}. In light-cone gauge we have a well defined Hamiltonian and time coordinate in string field theory \cite{light4, light5}. 
The Hamiltonian for string field theory in pp-background has also been constructed \cite{pp}. 
The string field theory in  pp-wave background is motivated  due to the  importance of Berenstein-Maldacena-Nastase (BMN) proposal \cite{casu4}. This background can be viewed as a deformation of the flat background. However, for pp-wave background,  the string theory is exactly solvable  in the light-cone gauge, even   in the presence of curvature and  form flux \cite{caus5, casu5a}. This makes the string theory on pp-wave background interesting, and we will use this background in this letter. It is non-trivial to understand causality in string field theory due to the extended nature of strings.   The causality  in the    free string field theory  has been investigated using  conditions for the   commutator of two open string fields to vanish \cite{casu1}. The causal structure in  string field theory  in light-cone guage has also been studied \cite{casu2}.  This was done by  treating the commutator as a distribution on infinite dimensional loop space. The causality  in the pp-wave light-cone gauge string field theory has also been discussed \cite{casu4, casu4a}. So, in this letter we  will analyze the causality for light-cone gauge string field theory on the pp-wave background.

We would like to point out that even for ordinary quantum field theory, it is non-trivial to properly investigate the causal structure of quantum processes. This is because in any  processes,   the initial value of a physical  quantity changes due to interactions. To obtain the information about such a  process, we have to obtain the information about this change. Now in a quantum process, such quantities are associated with eigenvalues of quantum operators. Thus, we have to measure the change in the eigenvalues of such quantum operators which can be done for ordinary quantum mechanical system by calculating the difference between the initial and final eigenvalues of such an operator.  The  procedure  to obtain the  difference between the initial and final   eigenvalues  is called the two-point measurement   scheme \cite{tasaki00,kurchan01}.   As in this scheme,   an   initial  projective measurements is made at the start of the quantum process. After making that initial measurement, the system is   perturbed, and the perturbation changes the eigenvalues associated with quantum operators. Then a final projective measurement is made to obtain the final eigenvalue. The difference in the eigenvalues is used to obtain the  distribution function. This can also be used to obtain the characteristic function for the system. 

This procedure cannot be directly used in the theories with Lorentz structure, as such projective measurements would  violate causal structure of the theory \cite{5cd,5mn, 6lk}. Hence, a new   scheme called the   Ramsey scheme  has been proposed to resolve these problems \cite{6ab, 6ba}.  
In this scheme, first an  auxiliary qubit is prepared in 
 a superposition of ground and excited states. This auxiliary qubit is coupled to the system with a Lorentz structure. Then   the information about the system is    transferred to the qubit. This  information can be obtained by applying Hadamard operator on the qubit. After that the system is perturbed, and the information about the change is also transferred to the qubit. This information is obtained by again  applying the Hadamard operator to  the qubit. Hence, the information about the process is obtained without making two projective measurements. 
The Ramsey scheme has been applied to quantum field theory, as most quantum field theories are constructed on spacetime with a Lorentz structure \cite{5}.   

In string theory   quantities, like mass and spin of elementary particles  are obtained  dynamically from  the conformal field theory of world-sheet   \cite{x1, x2}.   Different particles are  viewed as different vibrational modes of the same string, and the change in these modes can be seen as a change in the particle content of the system \cite{y0,y4}. This change occurs due to interaction with background fields \cite{y1, y2}. So, it is again possible to start by coupling the world-sheet of string theory to a suitable string auxiliary qubit, and then use the 
  string theoretical   Ramsey scheme to obtain information about such string theoretical processes \cite{x1, x2}. 
We note that the difference between the eigenvalues calculated from Ramsey scheme does not correspond to any quantum operator, as it represents a quantum process.
The information spreading in interacting string field theory has been related to causality \cite{spreading}. Here, we make the relation more rigorous by using techniques from quantum information theory based on quantum fisher information. 
It is known that physical quantities, which cannot be represented by a quantum   operator are   probed using  quantum fisher information   \cite{1a, 2a, 1b, 2b}.  Thus, such difference between the initial and final eigenvalues of a quantum operator can be probed using quantum fisher information. 
Such difference between the eigenvalues of the Hamiltonian has been used to analyze quantum work for such quantum processes   \cite{4a, 4b, 4c, 4d}.
In fact, quantum  fisher information has been to obtain investigate such  differences in  a string theoretical process \cite{x1, x2}. Here, we will generalize this work to a second quantized string theory, and explicitly analyze the quantum fisher information associated with a quantum process in string field theory. This will be done using light-cone gauge string field theory on   pp-waves background.

\section{Dyson Expansion in String Field Theory}
 
The study of  string field theory in  pp-wave background has been motivated from  the   BMN proposal \cite{casu4}. Even though this  background is  a deformation of the flat background, the    string theory is exactly solvable  in the light-cone gauge,  even  after such a deformation \cite{caus5, casu5a}. Furthermore,  the Hamiltonian for string field theory in pp-background has also been constructed \cite{pp}. As we have a well defined temporal coordinate along with this Hamiltonian we 
 will construct an analogous Dyson expansion for string field theory in pp-wave background. This can be done by first observing that 
the string theory in pp-wave background can be    studied using the following metric   \cite{casu4}
\begin{equation} \label{pp-wave metric}
ds^2 = -2 dx^+ dx^- - \mu^2 \sum_{i=1}^8 (x^i)^2 dx^+
dx^+ + \sum_{i=1}^8 dx^i dx^i,   
\end{equation} 
This background is viewed   as a deformation of the flat
background, with a curvature dependent  $\mu$ term \cite{caus5, casu5a}. 
So, in  light cone gauge, with  
$
x^+ = p^+ x^+ , 
$
the theory is exactly solvable, 
\cite{Kaku:zz}. Now in the  the light-cone gauge, the light-cone 
coordinates $x^-$ and the transverse coordinates  $x^i$ can be expressed as 
\begin{equation}
x^-(\sigma) = x^-_0 + \int d \tilde{\sigma} \;  {x^i}' P^i(\tilde{\sigma}) \\
x^i(\sigma) =  x^i_0 + \sqrt{2} \sum_{l=1}^\infty x^i_l \cos (l\sigma).
\label{mode open}
\end{equation}
The light-cone string field $\Phi = \Phi[x^+ , x_0^-, x^i(\sigma)]$ will depend on  $x^+, x_0^-, x^i(\sigma)$. So, it is possible to write a Schr\"{o}dinger  like equation for it as \cite{paper}
\begin{equation} 
\label{sch-eq1} i \frac{\partial \Phi}{\partial x^+}
=  \Tilde{H}_0 \Phi[x^+ , x_0^-, {x^i}(\sigma)].
\end{equation} 
Here,   $ \Tilde{H}_0$ denotes   the free light-cone
Hamiltonian (with $2\alpha'=1$) as 
\begin{equation}  \label{Hamiltonian}
 \tilde{H}_0 = \frac{\pi }{2 p^+} \int 
d\sigma \Bigg[ -\frac{\partial^2}{\partial {x^i}(\sigma)^2} + \frac{1}{\pi^2}
\Bigg( \frac{\partial x^i(\sigma)}{\partial \sigma}\Bigg)^2 + \frac{m^2}{\pi^2} (x^i(\sigma))^2
\Bigg] \quad \mbox{and} \quad m := \mu p^+. \end{equation}
It is possible to express $ \hat{H}_0$ is terms of modes as
\begin{equation} \label{sft eom modes}
i \frac{\partial \Phi}{\partial x^+} =
\frac{1}{2p^+} \sum_{l=0}^\infty \hat{H}_{l0} \ \Phi(x^+, x^-, \{x^i_l\}) ,
\end{equation} 
where  $
     \hat{H}_{l0} = - {\partial^2}/{\partial {x^i_l}^2} +\omega_l^2 (x^i_l)^2, $ and $ \omega_l =\sqrt{l^2 + (\mu p^+)^2}. 
$
This is the equation of a simple harmonic oscillator, with a well defined time coordinate. Using this expression, it is possible to write the string field as 
\begin{equation} 
\Phi = \int \frac{dp^+}{\sqrt{2\pi p^+}}
\sum_{\{n^i_l\}} A\left( p^+, \{n^i_l\} \right) e^{-i \left( x^+ p^- +
x^- p^+ \right)} \prod_{l=0}^\infty \varphi^l_{\{n^i_l\}} ( x^i_l ) +
h.c. 
\end{equation}
where 
\begin{equation} 
\varphi^l_{\{n^i_l\}} ( x^i_l )= \prod_{i=1}^{d-2}
H_{\{n^i_l\}} \left( \sqrt{\omega_l} x^i_l \right) e^{-\omega_l
({x^i_l})^2 / 2} \sqrt{ \frac{ \sqrt{\omega_l/\pi}}{2^{ n^i_l} (n^i_l!)}} .
\end{equation} 
The light-cone energy can now be defined as  
$
p^- = (p^+)^{-1}
 \sum_{i=1}^{d-2} \sum_{l=0}^\infty n^i_l \omega_l + z
$ with    $z$ a  the zero point energy. The conjugate momentum field can be expressed as 
see \cite{Kaku:zz}, 
\begin{equation} 
\Pi[x(\sigma)] = \frac{ \delta \mathcal{L} }{\delta \left(\partial\Phi / \partial x^+\right) } =   i \Phi. 
\end{equation}
Now we can write  the equal time commutation relations for the string field as 
\begin{equation} \label{ECTR} 
\left[ \Phi\left(x^+,x^-_0,\vec{x_l}\right),
\Phi\left(x^+,y^-_0,\vec{y_l}\right)\right] = \delta(x^-_0 - y^-_0)
\prod_{i=1}^{d-2} \delta\left[ x^i(\sigma) -
y^i(\sigma)\right]. 
\end{equation} 

Thus, it is possible to write the commutation relation  between the string
creation-annihilation operators as 
\begin{equation}
\left[ \hat{B}\left(p^+, \{n^i_l\} \right), \hat{B}^\dagger
\left( q^+, \{m^j_k\} \right) \right] = p^+ \delta(p^+ - q^+) \delta_{
\{n^i_l\},\{m^j_k\}} 
\label{cm relations} 
\end{equation}
We will denote these creation and annihilation states by a simpler index $\hat{B}^I$ and $\hat{B}^{1\dagger}$, where $I$ specifies the information relating to the $ (p^+, \{n^i_l\}) $. Thus, we can write the vacuum state as a state annihilated by $\hat{B}^I$, and $\hat{B}^{I \dagger}$ as the creation operator. 
Now we can construct a density state for string field theory in pp-waves as 
\begin{equation}
\hat{\rho}= \mathcal{N} \sum_I\sum_K d_K{d_I}^*|K\rangle\langle I|
\end{equation} 
Here, $ \mathcal{N}=1/\sum_K |d_K|^2$   is the normalization constant for string field theory. 
As $x^+$ acts as the time variable, we can view the evolution of   string field theory density state from   $\hat{\rho} (0)$ at time $x^+ =0$ to $\rho (t)$ at 
time $x^+ = T$
\begin{equation}
   \hat{\rho}(t)=\hat{U}(t)\hat{\rho} (0) \hat{U}^{\dagger}(t). 
\end{equation}
Here, $U$ is a suitable unitary operator, constructed from the Hamiltonian $H$ and time $x^+$.  Now we can add an interaction term to this Hamiltonian $H_I$, and write the total Hamiltonian as 
\begin{equation}
    \hat{H} = \Tilde{H}_0+\Tilde{H}_I,
\end{equation}
This interaction can be constructed using different polynomial functions $ \hat{H}_\mathcal{F} = \mathcal{F} [\Phi(x^+,x^-_0,\vec{x_l} )]$ of the sting field $\Phi (x^+,x^-_0,\vec{x_l} )$ with a suitable couple constant $\lambda$. We will also constraint the interactions, such that the interactions will start at a time $x^+ =0$ and end at a time $x^+ =T$. This can be achieved by a switching function $(x^+)$, such that $\chi(x^+) $    which turns the interaction $\mathcal{F}$   on at $x^+ >0$ and off at $x^+ < t$. Using $\chi(x^+) $, we can write the interaction term as  $\Tilde{H}_I  = \lambda\chi (x^+) \hat{H}_\mathcal{F}$. 
    
 As the  interaction is turned on at  $x^+ >0$ and turned off at  $x^+ = t$, the total Hamiltonian coincides  with the free Hamiltonian  at $\hat{H}(0)= \hat{H}_0$ and $\hat{H}(t)= \hat{H}_0$ because  $\hat{H}_I(0) =0$ and $\hat{H}_I(t) = 0 $.  So, using this interaction Hamiltonian $ \hat{H} $, we can write the unitary transformation $\hat{U}$ as 
      \begin{equation}
      \hat{U} (x^+) = \mathcal{T} \exp{\left(-i  \lambda\chi (x^+)   \Tilde{H}  \right)}, 
    \end{equation}
Here, the  time ordering between $0<x^+ <t$ is represented  by  $\mathcal{T}$.     Using analogous  Dyson expansion for string field theory in pp-waves, the   unitary operator can be expanded as 
    \begin{equation}
    \hat{U}(x^+ )=1+\hat{U}^1(x^+ )+\hat{U}^2(x^+ )+O(\lambda^3), 
    \end{equation}
    with $\hat{U}^1(x^+ )$ and  $\hat{U}^2(x^+ )$ representing  the first and second order terms in this analogous  Dyson expansion for string field theory.  
    The Dyson expansion was used to obtain the characteristic function for string theory in light-cone gauge \cite{x1, x2}. We will now use it for analyzing the characteristic function  for string field theory.

\section{String Field Theoretical Ramsey scheme }
In this section, we will develop a string field theoretical Ramsey scheme. 
The Ramsey scheme can be used to obtain information about a quantum process for a system  with a Lorentz structure on it \cite{5}. This is because two projective measurements   cannot be made on a system with a Lorentz structure \cite{5cd,5mn, 6lk}. This problem will occur in string field theory due to   Lorentz structure of target space, as string field theory is a field theory defined on it. So, we cannot make two projective measurements on string field, and have to develop a  string field theoretical Ramsey scheme to properly address the problem of causality. We start from a 
 a string field theoretical  operator $\hat{A}_I (0)$, with
 with eigenvalues $  A_I$ and eigenvectors $|A_I \rangle$.
 We could use the string field theoretical Hamiltonian $\hat{A} = \hat{H}$, but we will keep the analysis general. This string field theoretical operator is changed due to perturbations $\mathcal{F}$.  Then after some time $x^+ = t$, this interaction is turned off, and the operator $  \hat{A}(t)$ has eigenvalues $\tilde A_J$
and eigenvectors $|{\tilde A}_J\rangle $. To analyze this change in the eigenvalues, we generalize the Ramsey scheme \cite{6ab, 6ba} to string field theory on a pp-wave background. We first observe that the operator can be expressed as 
  \begin{eqnarray}
    \hat{A}(0)=\sum_{i}A_I|{A}_I\rangle \langle {A}_I|,   \,\,\,\,\,
    \,\,\,\,\,\,\,\,\,\,\,\,\,\,\,\,\,\,\,\,\,\,\,\,\,
    \hat{A}(t)=\sum_{J}\tilde A_J|{\tilde A}_J\rangle \langle {\tilde A}_J|.
    \end{eqnarray}
 We can represent the  probability to  measure eigenvalue $A_I$ at time $x^+ =0$  as $p^0_ I = |\langle A_I|\rho|A_I\rangle|^2$. The interaction produced by $\mathcal{F}$ changed the  eigenvectors of the operator from   $|A_I\rangle $ to $|\tilde A_l\rangle $. Now at time $x^+ = t$, we  can also write the  conditional probability for measuring eigenvalue $\tilde A_J$. Here, we have  assumed the initial eigenvalue is $A_I$ at  $x^+ =0$. So,  we can write   $p^{x^{+}}_{l|I} = \langle \tilde A_J|U | A_I\rangle|^2 $.  We can represent the  difference between   $A_I$ and $\tilde A_J$ as 
$
\Bar{A}_{I,J}= \tilde A_J-A_I.
$
Now as we have an expression for  $ p^0_I $ and $ p^{x^+ }_{l|I}$, we can write  the  probability to obtain   $\Bar{A}^{I,J}$ as     
  \begin{eqnarray}
{p_{I,J}} = p^0_I \  p^{x^+ }_{J|I} =  |\langle {A}_I|\rho|{A}_I\rangle|\langle \tilde {A}_J|U | {A}_I\rangle|^2.
\end{eqnarray}
Using these probabilities, we can obtain   the   probability distribution for  $\Bar{A}_ {I,J}$. To obtain this distribution, we can  define a new string field theoretical  distribution variable   $\mathcal{A}$. This variable   is  $\Bar{A}_{I,l}$, in the absence of degeneracies. Here, we have defined $\mathcal{A}$ to 
  account for degeneracies in the eigenvalues of the string field theoretical  operator. So, we can write the probability distribution for  $\mathcal{A}$ as $p(\mathcal{A})$, where 
\begin{equation}
p(\mathcal{A})=\sum_{IJ}p_{I,J}\delta(\mathcal{A}-\Bar{A}_{I,J}).
\end{equation}
The average value for $\mathcal{A}$ can be obtained using    $P(\mathcal{A})$ as 
  \begin{eqnarray}
           \Bar{ \mathcal{A}}=\int \sum_{IJ}p_{I,J}\delta(\mathcal{A}-\Bar{A}_{I,J})\mathcal{A}d\mathcal{A}.
     \end{eqnarray}
Here, we can use the expression for $p_{I,J}$, and  write $ \Bar{ \mathcal{A}}$ as
       \begin{eqnarray}
           \Bar{ \mathcal{A}}
         &=& \sum_{IJ} |\langle {A}_I|\rho|{A}_I\rangle|\langle \tilde {A}_J|U | {A}_I\rangle|^2 \left[\tilde A_J (x^+ ) -A_I (0)\right]. 
         \end{eqnarray}
Using      $ p_J (x^+ ) = \sum_I p _{I,J} (x^+ )$ and $p_I (0) = \sum_J p_{I,J} (0) $, we    observe that $ \Bar{ \mathcal{A}}$ can be written as 
 \begin{equation} 
 \Bar{ \mathcal{A}} = \sum_J p_J (x^+ ) A_l{(x^+ )}-\sum_I p_I (0) A_I{(0)},
 \end{equation}  
Furthermore, as  $\sum_J p_J (x^+ ) A_J{(x^+ )} = tr[\mathcal{A}{(x^+ )}\rho{(x^+ )}] $ and $\sum_I p_I (0) A_I{(0)}  = tr[\mathcal{A}{(0)}\rho{(0)}] $, we obtain  
\begin{equation}
  \Bar{ \mathcal{A}} = tr[\mathcal{A}{(x^+ )}\rho{(x^+ )}]-tr[\mathcal{A}{(0)}\rho{(0)}].
\end{equation}

The corresponding characteristic function of  $\mathcal{A}$ can be expressed as 
  \begin{equation}
    \Tilde{P}(\nu)=\int P(\mathcal{A})e^{i\nu \mathcal{A}} d\mathcal{A}=\langle e^{i\nu \mathcal{A}}\rangle.
  \end{equation} 
 The parameter $\nu$  is used  in the Ramsey scheme.
We start by defining  a string field theoretical auxiliary qubit, which can couple to the density matrix defined for a light-cone gauge string field theory. The excited state of such an  string field theoretical auxiliary qubit is represented by   $|1 \rangle$ and its ground state is represented by  $|0 \rangle$. Such a configuration of this  string field theoretical auxiliary qubit can be used to  transfer data from a string field theoretical density matrix to this qubit. So, we first couple it to the string field theoretical density matrix  $\rho (0)$ defined for string field theory on a pp-wave background. Both the string field and the qubit are in this ground state. Thus, the system can be expressed as a product state 
\begin{equation}
\hat{\rho}_\text{tot}=   \hat{\rho}\otimes\hat{\rho}_\text{aux}.
\end{equation}
We can construct a string field theoretical  Hadamard operator, which will operate on this string field theoretical   qubit. We apply this string field theoretical  Hadamard operator on  the auxiliary qubit after coupling it to the  the string field theoretical density state. 
After that we make the string field theory to interact, using the $\chi(x^+)$ function, and this is achieved using the   unitary operator $\hat{U}(x^+)$ constructed from $\hat{H}$ and time $x^+$. This can be done by generalizing the results from field theory   \cite{5}, to string field theory, and thus we obtain  
 \begin{equation}
     \hat{C} (\nu) = \hat{U} e^{-i\nu  \hat{H}(0) } \otimes|0\rangle\langle0|+ e^{-i\nu  \hat{H}(t)}\hat{U}\otimes|1\rangle\langle1|.
 \end{equation}
 The  string field theoretical qubit state can  be represented as
 \begin{equation}
    \hat{\rho}_\text{aux}= \text{Tr} _{X}[\hat{C}(\nu) \hat{\rho}_\text{tot} \hat{C}^\dagger (\nu)], 
\end{equation}
where $\text{Tr}_{X}$ is defined to be the      traces over the string field theoretical states. A final string theoretical Hadamard operation is performed on the string field theoretical   qubit state to obtain the information about this string field theoretical interaction $\mathcal{F}$.  Here, we have obtained an expression for $\hat {\rho}_ \text{aux}$. It will  depend on the   form of the operator $\hat{A}$ as well as the form of the interaction. A general expression for for such a state can be written as  \cite{6ab, 6ba} 
\begin{equation}
\hat{\rho}_\text{aux}    =
\frac12\Big\{\mathbb{1} + \text{Re} [\tilde{P}(\nu )]\hat{\sigma}_z + \text{Im} [\tilde{P}(\nu )]\hat{\sigma}_y\Big\}~. 
\label{eq:rho_aux}
\end{equation}
This general  expression  can be compared to the      string  field theoretical auxiliary qubit  $\hat{\rho}_ \text{aux}$ to obtain an expression for $\tilde{P}(\nu)$. 
Thus, we can explicitly obtain the  characteristic function for light-cone  string field theory, without violating causality. 
\section{Fisher Information}
The string field theoretical characteristic  function can be used to obtain the average difference for any string field theoretical operator $\hat{A}$. 
The interaction of string fields, and its effect on physical quantities can be analyzed using this string field theoretical characteristic function.  We now denote the 
average difference by $\Bar{\mathcal{A}}$. This represents the average of the difference between the initial and final eigenvalues of a string field theoretical operator $\hat{A}$. This average of the difference can be obtained from the characteristic  function by generalizing the field theoretical results \cite{5} to string field theory. 
So, we can express $\Bar{\mathcal{A}}$ as
 \begin{eqnarray}
  \Bar{ \mathcal{A}} =i  \frac{d}{d\nu}  
  \Tilde{P}(\nu).  
 \end{eqnarray}
Here, we start from initial states, and the system is perturbed by $\mathcal{F}$ and then we have the final states. The Ramsey scheme \cite{6ab, 6ba} is generalized to string field theory, and this string field theoretical Ramsey scheme can be  used to obtain the difference between the initial and final eigenvalues of an operator.
The we can represent $\Bar{\mathcal{A}}$ to be the average of such difference, and this can has been related to the characteristic function. The details of this average will also depend on the explicit form of the operator $\mathcal{A}$. We note that in any process, when 
 $\mathcal{A}$ does not contain  information about string field theoretical states, this average will disappear 
  $\Bar{\mathcal{A}}= 0$. Thus, it is important to analyze the information associated with $\mathcal{A}$. This can be done using the quantum information theoretical technique based on quantum fisher information. We also observe that when the operator is the Hamiltonian,  $\hat{A} = \hat{H}$, the   average of the difference between the initial and final eigenvalues of $\hat{H}$ represented by  $\Bar{\mathcal{H}}$ corresponds to quantum work distribution. The difference between the eigenvalues of the Hamiltonian has been used to analyze quantum work  distribution in different quantum systems  \cite{4a, 4b, 4c, 4d}. Quantum work distribution has been calculated for field theory using Ramsey scheme \cite{5}, and here we have generalized it for string field theory in light-cone gauge.

The quantum fisher information  is used to probe those physical quantities which cannot be related to operators  \cite {1a, 2a, 1b, 2b}. Here, we have observed that the difference between the initial and final eigenvalues of an operator in string field theoretical process cannot be represented by an quantum operator, and so we will use quantum fisher information to analyze it. The change in the physical quantity $\mathcal{A}$ can be related to  a parameter $\nu$. The change in the quantum states in a string field theoretical process can be probed by the change in the eigenvalues of $\mathcal{A}$ only if if depends on the states. Thus, dependence can be quantified using quantum fisher information. So, a change in the quantum fisher information will occur in a string field theoretical process, if the string field theoretical operator contains quantum fisher information about the string fields.    The quantum fisher information  associated with $\mathcal{A}$ can be expressed as
\begin{eqnarray}
   F(\nu) =\int P(\mathcal{A}) \left| \frac{\partial }{\partial \nu }\log P(\mathcal{A})\right|^2 d\mathcal{A}.
\end{eqnarray}
This can be related to the root mean square error in the estimation of $  \nu$ using the quantum  Cramer-Rao bound \cite{est1, est2}. Now  $\delta \nu$ the root mean square error in the estimation of $\nu$, then it is related to  the number of times the system has been probed $(\zeta)$ and quantum Fisher information $  F(\nu)$ as 
\begin{equation}
  \delta \nu \geq \frac{ 1}{ \sqrt{\zeta F(\nu)}}.   
\end{equation}
Now if a quantity does not depend on string fields,  the quantum fisher information associate with it vanishes, and the  root mean square error in the estimation of $  \nu$ diverges, and so we cannot get any information about $\nu$.
We first define a quantity by a quantum operator in string field theory. If this quantum operator $\mathcal{A}$ has quantum fisher information about string fields, then a change in the quantum state of those string fields, will also change the quantum fisher information associated with that operator. This can be used to analyze such a   string theoretical process. As the quantum fisher information is  not obtained through direct measurement, it will not violate causality. Thus, we are able to investigate such string field theoretical processes without breaking causality. 
  
\section{Conclusion}
In this letter, we have  investigated causality in string field theory. This was  done using  pp-wave light-cone gauge string field theory. As we had a well defined Hamiltonian and temporal coordinate for  light-cone gauge string field theory, we constructed a Dyson expansion for string field theory. This was then used to analyze a string theoretical process by   generalizing the  Ramsey scheme to string field theory. We first coupled the string field to a suitable qubit, and transformed the information about the string field to that qubit. Then we perturbed the string field, and transferred the information about the perturbation  to the qubit. Thus, we extracted the information  about the string theoretical process from that qubit.  We   explicitly constructed  a characteristic function for  interactive string field theory. This  characteristic function was  used to  obtain average of the difference between the initial and final values of such an operator defined in string field theory.  We finally  used the quantum information theoretical technique based on fisher information to extract the information about this string theoretical process. We obtained an explicit expression for quantum fisher information. The consequences of this quantum fisher information were investigated using the quantum  Cramer-Rao bound. 

It would be interesting to develop this formalism further and apply to specific string field theoretical processes, and explicitly demonstrate how causality is not violated in string field theoretical processes. This can also be generalized to covariant  string field theory. As the covariant string field theory can be related to light-cone gauge string field theory, we can first calculate a string field theoretical process  in light-cone gauge, and explicitly demonstrate how the causality is not violated in it. Then we can map that analysis to covariant gauge, and this way we can analyze causality in covariant gauge. 
We can also use quantum fisher information to investigate causality in covariant string field theory. It would also be interesting to repeat this analysis on other backgrounds, and analyze the violation of causality on those backgrounds. It would be interesting to generalize   Ramsey scheme for string field theory on such backgrounds. This can be used to obtain the characteristic function for string field theory on those backgrounds. Then causality in such string field theoretical processes can be investigated using quantum fisher information.  

In this letter, we analyzed the causality in light-cone guage. It will be interesting to analyzing the causality using Ramsey scheme in covariant gauge. It is possible to relate the partition functions in two different gauge using finite field BRST transformations \cite{ff12, ff14, ff16, ff18, ff19}. To construct a finite field BRST transformation, the   infinitesimal global parameter  in the BRST
  transformations is first made field dependent, and then   integrated to obtain a finite field dependent BRST transformations \cite{ff10, ff15}. The finite field BRST transformation for string theory has already been constructed \cite{ff20, ff22}. It would be interesting to use the finite field BRST transformation along with the results of this letter to investigate the causality in covariant gauge.

\section*{Acknowledgments}
Salman Sajad Wani acknowledges support from the Scientific and Technological Research Council of Türkiye (TÜBİTAK) BİDEB 2232-A program under project number 121C067.


\begin{thebibliography}{100}
\bibitem{witten} E. Witten,  Nucl.Phys.B  {268},  270 (1986) 

 \bibitem{shell1}V.~Forini, G.~Grignani and G.~Nardelli, JHEP {04}, 053 (2006) 
\bibitem{shell2} B.~Urosevic, Phys. Rev. D {50}, 4075 (1994) 
  \bibitem{back1}A.~Sen and B.~Zwiebach, Nucl. Phys. B  {414}, 649 (1994)  
  \bibitem{back2}A.~Sen, JHEP {02}, 155 (2018)  


\bibitem{sft01}A.~Sen, JHEP {12}, 075 (2015) 
\bibitem{sft02} A.~Sen, JHEP {11}, 050 (2016)  
  \bibitem{prob6}A.~Sen, JHEP {12},  115 (016)    \bibitem{prob7} C.~De Lacroix, H.~Erbin and A.~Sen, JHEP  {05},  139 (2019)  


 \bibitem{light}P.~Lee, S.~Moriyama and J.~w.~Park,
Phys. Rev. D {67}, 086001 (2003)
\bibitem{light2}H.~Matsunaga, JHEP {04}, 143 (2019)
\bibitem{light4}G.~Siopsis, Phys. Lett. B  {195}, 541 (1987)
\bibitem{light5}C.~Batlle and J.~Gomis, Phys. Lett. B  {187}, 61-66 (1987)
\bibitem{pp} J.~Gomis, S.~Moriyama and J.~w.~Park, Nucl. Phys. B  {659}, 179-192 (2003)


 \bibitem{x11} S.~Hellerman and I.~Swanson, Phys. Rev. Lett. {114} 111601 (2015)
 \bibitem{x21} F.~Rojas and C.~B.~Thorn, Phys. Rev. D {84}, 026006 (2011)
 \bibitem{y0}A.~E.~Lawrence and E.~J.~Martinec, Class. Quant. Grav.  {13}, 63-96 (1996)
 \bibitem{y4}I.~Bars and K.~Sfetsos, Phys. Rev. D {46}, 4510-4519 (1992)
 \bibitem{y1} J.~T.~Liu and R.~Minasian, Nucl. Phys. B  {874}, 413-470 (2013)
\bibitem{y2} M.~Schnabl, JHEP {11}, 031 (2000)

 \bibitem{tasaki00} S. Suomela, P. Solinas, J. P. Pekola, J. Ankerhold and  T. Ala-Nissila, Phys. Rev. B 90, 094304 (2014)

\bibitem{kurchan01}  F. W. J. Hekking and J. P. Pekola, Phys. Rev. Lett. 111, 093602 (2013)
\bibitem{5cd} M. Redhead, Found. Phys. 25, 123 (1995)
\bibitem{5mn}D. M. T. Benincasa, L. Borsten, M. Buck, and
F. Dowker, Class. Quantum Grav. 31, 075007 (2014)
\bibitem{6lk}R. D. Sorkin, gr-qc/9302018 (1993)


\bibitem{6ab}R. Dorner, S. R. Clark, L. Heaney, R. Fazio, J. Goold
and V. Vedral, Phys. Rev. Lett. 110, 230601 (2013)
\bibitem{6ba}L. Mazzola, G. D. Chiara and M. Paternostro, Int. J.
Quantum Inf. 12, 1461007 (2014)
 \bibitem{5}A.~Ortega, E.~McKay, \'A.~M.~Alhambra and E.~Mart\'\i{}n-Mart\'\i{}nez, Phys. Rev. Lett.  {122}, 24, 240604 (2019)
 \bibitem{x1}S.~S.~Wani, A.~Shabir, M.~Faizal and S.~Rubab,
EPL {139}, no.4, 42002 (2022) 
 \bibitem{x2} S.~S.~Wani, J.~Q.~Quach and M.~Faizal,
EPL {139}, no.6, 62002 (2022)

 \bibitem{spreading}D. A. Lowe, L. Susskind and J. Uglum,  Phys. Lett. B 327, 226 (1994) 
\bibitem{1a}J. Liu, H. Yuan, X-M. Lu and X. Wang, J. Phys. A. Math. Theor. 53, 023001 (2020)

\bibitem{2a}A. T. Rezakhani, M. Hassani and S. Alipour, Phys. Rev. Lett. 109, 190404 (2012)

\bibitem{1b}Y. Yang, Phys. Rev. Lett. 123, 110501 (2019).
\bibitem{2b}B. M. Escher, L. Davidovich, N. Zagury and R. L. de Matos Filho
Phys. Rev. Lett. 109, 190404 (2012)




 



\bibitem{Kaku:zz} M.~Kaku and K.~Kikkawa,  Phys.\ Rev.\ D { 10}, 1110 (1974) 

 
 


\bibitem{4a}A. del Campo, I. L. Egusquiza, M. B. Plenio and S. F. Huelga
Phys. Rev. Lett. 110, 050403 (2013)
\bibitem{4b}Erik Lucero, M. Hofheinz, M. Ansmann, Radoslaw C. Bialczak, N. Katz, Matthew Neeley, A. D. O’Connell, H. Wang, A. N. Cleland and John M. Martinis
Phys. Rev. Lett. 100, 247001 (2008)
\bibitem{4c}L. Rigovacca, A. Farace, L. A. M. Souza, A. De Pasquale, V. Giovannetti and G. Adesso, Phys. Rev. A 95, 052331 (2017)
\bibitem{4d}R. Nichols, P. Liuzzo-Scorpo, P. A. Knott and G. Adesso, Phys. Rev. A 98, 012114  (2018)

\bibitem{paper} Chu, Chong-Sun and Kyritsis, Konstantinos, Phys. Lett. B 566,  240 (2003)


\bibitem{casu1} H.~Hata and H.~Oda, Phys. Lett. B  {394}, 307-314 (1997)
\bibitem{casu2} D.~A.~Lowe, Phys. Lett. B  {326}, 223-230 (1994)
 \bibitem{casu4} D. Berenstein, J. M. Maldacena and H. Nastase,   JHEP 0204, 013 (2002)
 \bibitem{casu4a} C.~S.~Chu, Fortsch. Phys.  {53}, 436-441 (2005)
 \bibitem{caus5} R. R. Metsaev and A. A. Tseytlin,   Phys. Rev. D 65, 126004 (2002) 
 \bibitem{casu5a} R. R. Metsaev,  Nucl. Phys. B 625, 70 (2002)
\bibitem{est1}S. L. Braunstein and C. M. Caves, 
Phys. Rev. Lett. 72, 3439 (1994)
\bibitem{est2}S. L. Braunstein, C.  M. Caves and G. J. Milburn, Annals  Phys. 247, 135 (1996)

\bibitem{ff12}S. D. Joglekar and B. P. Mandal, Phys. Rev. D51, 1919 (1995)
  \bibitem{ff14}S.~Upadhyay, Phys. Lett. B {740}, 341-344 (2015)
 \bibitem{ff16}S.~Upadhyay, S.~K.~Rai and B.~P.~Mandal, J. Math. Phys.  {52}, 022301 (2011)
 \bibitem{ff18}S.~Upadhyay and B.~P.~Mandal, Annals Phys.  {327}, 2885 (2012)
 \bibitem{ff19} S.~Upadhyay, Annals Phys.  {356}, 299 (2015)
 
 \bibitem{ff10}S.~Upadhyay and B.~Paul, Eur. Phys. J. C {76},  394 (2016)
 

 \bibitem{ff15} M.~Faizal, B.~P.~Mandal and S.~Upadhyay, Phys. Lett. B {721}, 159-163 (2013)
 \bibitem{ff20}V.~K.~Pandey and B.~P.~Mandal,
EPL  {122},   21002 (2018)
  \bibitem{ff22}V.~K.~Pandey and B.~P.~Mandal, EPL  {125},   21001 (2019)



  



\end{thebibliography}
\end{document}